\documentclass[a4paper,12pt]{article}
\usepackage{epsfig}
\usepackage{axodraw}
\usepackage{graphicx}

\newlength{\dinwidth}
\newlength{\dinmargin}
\begin{document}
\title{$B$ to light meson transition form factors calculated  in perturbative QCD approach}
\author{Cai-Dian L\"u    and Mao-Zhi Yang
 \thanks{Email address: lucd@mail.ihep.ac.cn (C.D.L\"u), yangmz@mail.ihep.ac.cn (M.Z.Yang)}\\
{\small CCAST (World Laboratory), P.O. Box 8730, Beijing 100080, China;}\\
{\small Institute of High Energy Physics,
 P.O. Box 918(4), Beijing 100039, China}\footnote{Mailing address}
}

\maketitle
\begin{picture}(0,0)
       \put(365,270){BIHEP-TH-2002-62}
       \put(365,250){\bf hep-ph/0212373}
\end{picture}

\begin{abstract}

We calculate the $B\to P$, $B\to V$ (P: light pseudoscalar meson,
V: light vector meson) form factors in the large-recoil limit
in perturbative QCD approach, including both the vector (axial vector) and tensor
operators. In general there are two leading components $\phi_B$
and $\bar{\phi}_B$ for $B$ meson wave functions. We consider both
contributions of them. Sudakov effects ($k_{\perp}$ and
threshold resummation) are included to regulate the soft
end-point singularity. By choosing the hard scale as the maximum
virtualities of internal particles in the hard $b$-quark decay
amplitudes, Sudakov factors can effectively suppress the
long-distance soft contribution. Hard contribution can be dominant
in these approaches.

\end{abstract}

\bigskip


\newpage

\section{Introduction}
The most difficult task in calculating $B$ meson decay amplitude
is to treat the hadronic matrix element $\langle M_1
M_2|Q_i|B\rangle$, which is generally controlled by soft
non-perturbative dynamics of QCD. Here $Q_i$ is one of the effective
low energy transition operators of $b$ quark decays \cite{buras},
$M_1$ and $M_2$ are the final state mesons produced in $B$ decays. In
the earlier years, this hadronic matrix elements of $B$ decays
were treated by an approximate method, which is called
factorization approach \cite{factor}. In factorization approach
the hadronic matrix element of four-fermion operator is
approximated as a product of the matrix elements of two currents,
$
\langle M_1 M_2|Q_i|B\rangle \simeq \langle M_1|j_{1\mu}|0\rangle
 \langle M_2|j_2^{\mu}|B\rangle ,
$
where $j_{1\mu}$ and $j_2^{\mu}$ are the two relevant currents
which can be related to $Q_i$ through $Q_i=j_{1\mu} j_2^{\mu}$.
The matrix element of $j_{1\mu}$ sandwiched between the
vacuum and meson state $M_1$ directly defines
the decay constant of $M_1$. For example, if $M_1$ is a pseudoscalar and
$j_{1\mu}$ is $V-A$ current, the relation between the matrix element and
the decay constant will be $\langle M_1|j_{1\mu}| 0\rangle =if_{M_1}
p_{\mu}$, where $f_{M_1}$ and $p_{\mu}$ are the decay constant and
the four-momentum of $M_1$, respectively. The other matrix element
$\langle M_2|j_2^{\mu}|B\rangle$ can be generally decomposed into
transition form factors of $B\to M_2$ due to its Lorentz property.
The explicit definition of $B$ meson transition form factors
through such matrix elements can be found in section 4 and 5.

In semi-leptonic decays of $B$ meson, the decay amplitude can be
directly related to $B$ meson transition form factors without
 the factorization approximation. For example, for
$B\to \pi \ell \bar{\nu_\ell}$, the decay amplitude can be written
in the form:
\begin{equation}
{\cal A}(p_B,p_\pi)=\frac{G_F}{\sqrt{2}}V_{ub}(\bar{\ell}\gamma_{\mu}
 (1-\gamma_5)\nu_{\ell})\langle \pi (p_\pi)|\bar{u} \gamma^\mu
 b|\bar{B}(p_B)\rangle,
\end{equation}
where the form factors $F_1(q^2)$ and $F_0(q^2)$ are defined through the $B\to \pi$
transition matrix element $\langle \pi (p_\pi)|\bar{u} \gamma^\mu
b|\bar{B}(p_B)\rangle$ in
 eq.(\ref{5}) of section 4.  In general the
form factors are functions of momentum transfer squared
$q^2=(p_B-p_\pi)^2$. In the region of small recoil,
where $q^2$ is large and/or the final particle is heavy enough,
the form factors are dominated by soft dynamics, which is out of
control of perturbative QCD. However, in the large recoil region
where $q^2\to 0$, and when the final particle is light (such as the pion), $5
\mbox{GeV}$ $(m_B=5 \mbox{GeV})$ of energy is released. About half
of the energy is taken by the light final particle, which suggests
that large momentum is transferred in this process and the
interaction is mainly short-distance. Therefore perturbative QCD
can be applied to $B$ to light meson transition form factors in
large recoil region.

 Before applying perturbative method in this
calculation, one must separate soft dynamics from hard interactions.
This is called factorization in QCD.
The factorization theorem has been worked out
in ref.\cite{li} based on the earlier works on the applications of
perturbative QCD in hard exclusive processes \cite{hard}, where the
soft contributions are factorized into wave functions or
distribution amplitudes of mesons, and the hard part is treated
by perturbative QCD. Sudakov resummation has been introduced to
suppress the long-distance contributions. Recently this approach
has been well developed and extensively used to analyze $B$ decays
\cite{keum,luy,chenli,kurimoto,kurilisanda,lu,kou,chenkl,smishima}.
There is also another direction to prove   factorization in the
soft-collinear effective theory \cite{dan}, which shows correctly
the power counting rules in QCD.
In this work we shall calculate a set of $B\to P$ and $B\to V$
(P: light pseudoscalar meson, V light vector meson) transition
form factors in perturbative QCD (PQCD) approach. We use the $B$
wave functions derived in the heavy quark limit recently
\cite{kkqt}, and include Sudakov effects from transverse momentum
$k_\perp$ and threshold resummation \cite{kurilisanda,hnli}. In
general there are two Lorentz structures for $B$ wave functions.
If they are appropriately defined, only one combination gives
large contribution, the other combination  contributes 30\%.

A direct calculation of the
one-gluon-exchange diagram for the $B$ meson transition form
factors suffers singularities from the end-point region of the
light cone  distribution amplitude with a
momentum fraction $x\to 0$ in longitudinal direction. In fact, in the
end-point region the parton transverse momenta $k_\perp$ are not
negligible. After including parton transverse momenta, large double
logarithmic corrections $\alpha_s ln^2 k_\perp$ appear in higher
order radiative corrections and have to be summed to all orders.
In addition to the double logarithm like $\alpha_s ln^2 k_\perp$,
there are also large logarithms $\alpha_s ln^2x$ which should also
be summed to all orders. This is called threshold resummation \cite{hnli}.
The relevant Sudakov factors from both $k_\perp$ and threshold
resummation can cure the end-point singularity which makes the
calculation of the hard amplitudes infrared safe. We check
the perturbative behavior in the calculation of the $B$ meson
transition form factors and find that with the hard scale
appropriately chosen, Sudakov effects can effectively suppress
the soft dynamics, and the main contribution comes from the
perturbative region.

The content of this paper is as follows.  Section 2 is the
kinematics and the framework of the PQCD approach used in the calculation
of $B\to P$ and $B\to V$ transition form factors. Section 3
includes wave functions of $B$ meson and the light pseudoscalar
and vector mesons. We give the
results of $B\to P$ transition form factors in section 4, and
the $B\to V$ transition form factors in section 5. Section 6 are the numerical
results and discussion. Finally section 7 is a brief summary.

\section{The Framework}

Here we first give our conventions on kinematics.
In light-cone coordinate, the momentum is taken in the form
$k=(\frac{k^+}{\sqrt 2}, \frac{k^-}{\sqrt 2}, \vec k_{\bot})$ with
$k^{\pm}=k^0\pm k^3$ and $\vec k_{\bot}=(k^1, k^2)$. The scalar product of two
arbitrary vectors $A$ and $B$ is $A\cdot B=A_{\mu}B^{\mu}=
\frac{1}{ 2}(A^+B^- + A^-B^+) - \vec A_{\bot}\cdot \vec B_{\bot}$.
Our study is in the rest frame of B
meson. The mass difference of $b$ quark and $B$ meson is negligible in the heavy
quark limit and we take $m_b\simeq m_B$ in our calculation. The masses
of light quarks $u$, $d$, $s$ and light pseudoscalar mesons are
neglected, while the masses of light vector mesons $\rho$,
$\omega$, $K^*$ are kept in the first order.
The momentum of light meson  is chosen in the $``+"$ direction. Under these
conventions, the momentum of $B$ meson is
$P_B    =\frac{1}{\sqrt 2}( {m_B} ,  {m_B},\vec 0_\bot)$,
and in the large recoil limit $q^2\to 0$, the momentum of the light
pseudoscalar meson is $P_{P}=(\frac{m_B}{\sqrt 2}, 0, \vec
0_\bot)$. For the case of light vector meson, its momentum is
$P_V=\frac{m_B}{\sqrt{2}}(1,r_V^2,\vec 0_\bot)$ with $r_V$ defined
as $r_V\equiv m_V/m_B$. The longitudinal polarization of the vector meson is
$\varepsilon_L=\frac{1}{\sqrt{2}}\left (\frac{1}{r_V},-r_V,\vec
0_\bot \right )$, its transverse polarization
$\varepsilon_T=(0,0, \vec 1_\bot)$. The light spectator momenta $k_1$ in the
$B$ meson and $k_2$ in the light meson are parameterized as
$k_1=(0,x_1 \frac{m_B}{\sqrt{2}},k_{1\bot})$ and
$k_2=(x_2 \frac{m_B}{\sqrt{2}},0,k_{2\bot})$, where $k_2^-$ is
dropped because of its smallness (In the meson moving along the
`plus' direction with large momentum, the minus component of
its parton's momentum $k_2^-$ should be very small). We also
dropped $k^+_1$ because it vanishes in the hard amplitudes,
which can be simply shown below.

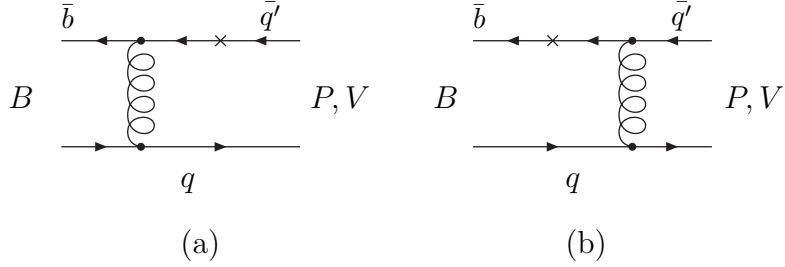
\begin{figure}[htbp]
 \begin{center}\scalebox{1.}{
 {
   \begin{picture}(140,120)(-30,0)
    \ArrowLine(30,60)(0,60)
    \ArrowLine(60,60)(30,60)
    \ArrowLine(90,60)(60,60)
    \ArrowLine(30,20)(90,20)
    \ArrowLine(0,20)(30,20)
    \Gluon(30,20)(30,60){5}{4} \Vertex(30,20){1.5} \Vertex(30,60){1.5}
    \Line(58,62)(62,58)
    \Line(58,58)(62,62)
    \put(-20,35){$B$}
    \put(94,35){$P,V$}
    \put(45,48){\small{}}
    \put(75,66){\small{$\bar{q'}$}}
    \put(0,65){\small{$\bar{b}$}}
    \put(45,5){$q$}
    \put(45,-20){(a)}
 \end{picture}
 }}
 \scalebox{1.}{
 {
   \begin{picture}(140,120)(-30,0)
      \ArrowLine(30,60)(0,60)
      \ArrowLine(60,60)(30,60)
      \ArrowLine(90,60)(60,60)
      \ArrowLine(60,20)(90,20)
      \ArrowLine(0,20)(60,20)
      \Gluon(60,20)(60,60){5}{4} \Vertex(60,20){1.5} \Vertex(60,60){1.5}
      \Line(28,62)(32,58)
      \Line(28,58)(32,62)
      \put(-15,35){$B$}
      \put(95,35){$P,V$}
      \put(15,48){\small{}}
      \put(75,66){\small{$\bar{q'}$}}
      \put(0,65){\small{$\bar{b}$}}
      \put(35,5){$q$}
      \put(35,-20){(b)}
 \end{picture}
 }}         \end{center}
 \caption{Diagrams contributing to the $B\to P,V$
form factors, where the cross denoting an appropriate gamma matrix.}
\end{figure}

The lowest-order diagrams for $B$ to light meson transition form
factors are displayed in Fig.1. The hard amplitude $H$ are
proportional to the propagator of the gluon, i.e.,
$H\propto 1/(k_2-k_1)^2 \simeq 1/(2k_2\cdot k_1)\simeq 1/(k_2^+k_1^-)$. It is
obvious that only $k_1^-$ left in the hard amplitude.

Factorization is one of the most important part of applying
perturbative QCD in hard exclusive processes, which separates
long-distance dynamics from short-distance dynamics.
The factorization formula for $B\to P,V$ transition matrix
element can be written as
\begin{eqnarray}
&\langle P,V(P_2)|\bar{b}\Gamma_{\mu} q^{\prime}|B(p_1)\rangle
=\displaystyle\int
dx_1dx_2d^2k_{1\bot}d^2k_{2\bot}\frac{dz^+d^2z_{\bot}}{(2\pi)^3}
\frac{dy^+d^2y_{\bot}}{(2\pi)^3}\nonumber\\
 &~~~~\times e^{-ik_2\cdot y}
 \langle P,V(P_2)|\bar{q}(y)_{\alpha}q^{\prime}_{\beta}(0)|0\rangle
 H^{\beta\alpha;\sigma\rho}_{\mu}  e^{ik_1\cdot z}
 \langle 0|\bar{b}(0)_{\rho} q_{\sigma}(z)|B(P_1)\rangle ,
 \label{factf}
\end{eqnarray}
where the matrix elements $\langle
P,V(P_2)|\bar{q}(y)_{\alpha}q^{\prime}_{\beta}(0)|0\rangle$ and
$\langle 0|\bar{b}(0)_{\rho} q_{\sigma}(z)|B(P_1)\rangle$ define
the wave functions of light pseudoscalar (vector) meson and $B$
meson, which absorb all the soft dynamics. $
H^{\beta\alpha;\sigma\rho}_{\mu}$ denotes the hard amplitude,
which can be treated by perturbative QCD. $\beta$, $\alpha$, $\sigma$
and $\rho$ are Dirac spinor indices. Both the wave functions and
the hard amplitude $H$ are scale dependent. This scale is usually called
factorization scale. Above this scale, the
interaction is controlled by hard dynamics, while the interaction
below this scale is controlled by soft dynamics, which is
absorbed into wave functions. The factorization scale is usually taken
the same as renormalization scale. In practice it is convenient
to work in transverse separation coordinate space ($b$-space)
rather than the transverse momentum space ($k_\bot$-space).
So we shall make a Fourier transformation $\int d^2k_\bot
e^{-i{\bf k}_\bot \cdot {\bf b}} $ to transform the wave functions
and hard amplitude into $b$-space. $1/b$ will appear as a typical
factorization scale.  As the scale $\mu>1/b$, the interactions are
controlled by hard dynamics, and as $\mu<1/b$ soft dynamics
dominates which is absorbed into wave functions.

Higher order radiative corrections to wave functions and
hard amplitudes generate large double logarithms through the
overlap of collinear and soft divergence. The infrared divergence
is absorbed into wave functions. The double logarithms
$\alpha_s ln^2 Pb$ have been summed to all orders to give
an exponential Sudakov factor $e^{-S(x,b,P)}$, here $P$ is the
typical momentum transferred in the relevant process, $x$ is
the longitudinal  momentum fraction carried by the relevant parton. The
resummation procedure has been analyzed and the result has been
given in \cite{li}, we do not repeat it here.

In addition to double logarithms $\alpha_s ln^2Pb$ in $b$-space (or say
$k_\bot$-space equivalently), radiative corrections to hard
amplitudes also produce large logarithms $\alpha_s ln^2x$.
These double logarithms should also be summed to all orders.
This threshold resummation leads to \cite{kurimoto}
\begin{equation}
S_t(x)=\frac{2^{1+2c} \Gamma (3/2+c)}{\sqrt{\pi} \Gamma(1+c)}
[x(1-x)]^c,
\end{equation}
where the parameter $c=0.3$. This function is normalized to unity.
$S_t(x)$ vanishes very fast
at the end-point region $x\to 0$ and $x\to 1$. Therefore
the factors $S_t(x_1)$ and $S_t(x_2)$ suppress the end-point
region of meson distribution amplitudes.

\section{The Wave Functions}

In the resummation procedures, the $B$ meson is treated as a heavy-light
system. In general, the B meson light-cone matrix element can be
decomposed as \cite{grozin,bene}
\begin{eqnarray}
&&\int_0^1\frac{d^4z}{(2\pi)^4}e^{i\bf{k_1}\cdot z}
   \langle 0|\bar{q}_\alpha(z)b_\beta(0)|\bar B(p_B)\rangle
   \nonumber\\
&=&\frac{i}{\sqrt{2N_c}}\left\{(\not p_B+m_B)\gamma_5
\left[\frac{\not v }{\sqrt{2}}\phi_B^+ ({\bf k_1})+\frac{ \not n}{\sqrt{2}}
{\phi}_B^-({\bf k_1})\right]\right\}_{\beta\alpha}
   \nonumber\\
&=&-\frac{i}{\sqrt{2N_c}}\left\{(\not p_B+m_B)\gamma_5
\left[\phi_B ({\bf k_1})+\frac{ \not n}{\sqrt{2}}
\bar{\phi}_B({\bf k_1})\right]\right\}_{\beta\alpha},
\label{aa1}
\end{eqnarray}
where $n=(1,0,{\bf 0_T})$, and $v=(0,1,{\bf 0_T})$ are the unit vectors
pointing to the plus and minus directions, respectively.
 From the above
equation, one can see that there are two Lorentz structures in the B meson
wave function.
In general, one should consider both these two Lorentz structures in
calculations of $B$ meson decays. The light cone distribution amplitudes $\phi^+_B$ and $\phi^-_B$
are derived by Kawamura et al. in
the heavy quark limit \cite{kkqt},
\begin{eqnarray}
\phi_B^+(x,b) &=&  \frac{f_B x}{\sqrt{6}  \Lambda_0^2} \theta (\Lambda_0-x)
J_0 \left[m_B b\sqrt{x(\Lambda_0-x}) \right]
 , \nonumber    \\
 \phi_B^-(x,b) &=&  \frac{f_B (\Lambda_0- x)}{\sqrt{6}  \Lambda_0^2} \theta (\Lambda_0-x)
 J_0 \left[m_B b\sqrt{x(\Lambda_0-x}) \right]
  ,
 \label{phib}
\end{eqnarray}
with  $\Lambda_0=2\bar\Lambda/M_B$, and $\bar \Lambda$ is a free parameter which should
be at the order of $m_B-m_b$.

The relation between $\phi_B$, $\bar \phi_B$ and $\phi_B^+$,
$\phi_B^-$ are
\begin{equation}
  \phi_B =\phi_B^+~,~~~~~~~~~~ \bar \phi_B   = \phi_B^+ -\phi_B^-  .
  \label{phiB}
\end{equation}
 The normalization conditions for these
two distribution  amplitudes are:
\begin{equation}
\int d^4 k_1\phi_B({\bf k_1})=\frac{f_B}{2\sqrt{2N_c}},
~~~\int d^4 k_1 \bar{\phi}_B({\bf k_1})=0.
\end{equation}
 From eqs.(\ref{phib}) and (\ref{phiB}), we can see that when
$x\to 0$, $\bar \phi_B \not\to 0$, while $ \phi_B \to 0$.
The behavior of $\phi_B$ with the definition (\ref{phiB}) is similar to the one defined in
previous PQCD calculations
\cite{keum,luy,chenli,kurimoto,kurilisanda,lu,kou,chenkl,smishima}.
Note that our definition of  $\phi_B$, $\bar \phi_B$ are
different from the previous one in the literature \cite{kurimoto}
\begin{eqnarray}
&&\int_0^1\frac{d^4z}{(2\pi)^4}e^{i\bf{k_1}\cdot z}
   \langle 0|\bar{q}_\alpha(z)b_\beta(0)|\bar B(p_B)\rangle
   \nonumber\\
&=&\frac{i}{\sqrt{2N_c}}\left\{(\not p_B+m_B)\gamma_5
\left[\frac{\not v }{\sqrt{2}}\phi_B^+ ({\bf k_1})+\frac{ \not n}{\sqrt{2}}
{\phi}_B^-({\bf k_1})\right]\right\}_{\beta\alpha}
   \nonumber\\
&=&-\frac{i}{\sqrt{2N_c}}\left\{(\not p_B+m_B)\gamma_5
\left[\phi_B' ({\bf k_1})+\frac{ \not n-\not v}{\sqrt{2}}
\bar{\phi}_B'({\bf k_1})\right]\right\}_{\beta\alpha},
\label{wb1}
\end{eqnarray}
with
\begin{equation}
  \phi_B' =\frac{\phi_B^+ +\phi_B^-}{2}~,~~~~~~~~~~
  \bar \phi_B'   =\frac{\phi_B^+ -\phi_B^-}{2}  .
  \label{wb2}
\end{equation}
This definition is equivalent to eqs.(\ref{aa1}, \ref{phiB}) in the
total amplitude.
 Although the final numerical
results should be the same, the form factor formulas are simpler using our new
definition.
Another outcome is that our new formula shows
explicitly the importance of the  leading twist contribution $\phi_B$,
which will be shown later.
However, if $\phi_B'$ and $\bar{\phi}_B'$ are defined    as
in eqs.(\ref{wb2}), both of their contributions  are equivalently
important (see the numerical results in Table 2 of
Ref.\cite{weiyang}). It is easy to check that both $\phi_B'$ and
$\bar \phi_B'$ here have non-zero endpoint at $x\to 0$. In this case, $\phi_B'$
does not correspond to the one defined in the previous PQCD
calculations \cite{kurimoto,kurilisanda}, where $\phi_B \to 0$, at endpoint,
when $x\to 0$ or
$1$.

The $\pi$, $K$ mesons are treated as a light-light system.
At the $B$ meson rest frame,
the $K$ meson (or pion) is moving very fast,  one  of $k_1^+$ or
$k_1^-$ is zero depending on the definition of the z axis.
We consider a kaon (or $\pi$ meson) moving in the plus direction in this
paper. The K meson distribution amplitude is defined by
\cite{ball}
\begin{eqnarray}
&&<K^-(P)|\bar{s_{\alpha}}(z)u_{\beta}(0)|0>\nonumber\\
&=&\frac{i}{\sqrt{2N_c}}\int_0^1 dx
e^{ixP\cdot z}\left[\gamma_5\not
P\phi_K(x) +m_0\gamma_5\phi_P(x)
-m_0\sigma^{\mu\nu}\gamma_5 P_{\mu}z_{\nu}
\frac{\phi_{\sigma}(x)}{6}\right]_{\beta\alpha}
\label{aa2}
\end{eqnarray}
For the first and second terms in the above equation,
we can easily get the projector of the distribution
amplitude in the momentum space. However, for the
third term we should make some effort to transfer it
into the momentum space. By using integration by parts
for the third term, after a few steps, eq.(\ref{aa2})
can be finally changed to be
\begin{eqnarray}
&&<K^-(P)|\bar{s_{\alpha}}(z)u_{\beta}(0)|0>\nonumber\\
&=&\frac{i}{\sqrt{2N_c}}\int_0^1 dx
e^{ixP\cdot z}\left[\gamma_5\not
P\phi_K(x) +m_0\gamma_5\phi_P(x)
+m_0[\gamma_5(\not n\not v-1)]\phi_K^t(x)\right]_{\beta\alpha}
\end{eqnarray}
where $\phi_K^t(x)=\frac{1}{6}\frac{d}{dx}\phi_{\sigma}(x)$,
and vector $n$ is parallel to the $K$ meson momentum $p_K$.
 And $m_{0K}=m_K^2/(m_u+m_s)$
 is a scale characterized by the Chiral
perturbation theory. For $\pi$ meson, the corresponding  scale is
defined as  $m_{0\pi}=m_\pi^2/(m_u+m_d)$.

For the light vector meson $\rho$, $\omega$ and $K^*$, we need distinguish
their longitudinal polarization and transverse polarization. If the ${K^*}$ meson
(so as to other vector mesons) is
 longitudinally polarized, we can write  its wave function
in longitudinal polarization \cite{kurimoto,ball2}
\begin{eqnarray}
&&<{K^*}^-(P,\epsilon_L)|\bar{d_{\alpha}}(z)u_{\beta}(0)|0>\nonumber\\
&=&
\frac{1}{\sqrt{2N_c}}\int_0^1 dx e^{ixP\cdot z}
\left\{ \not \epsilon \left[\not p_{K^*}
\phi_{K^*}^t (x) + m_{K^*} \phi_{K^*} (x) \right]
+m_{K^*} \phi_{K^*}^s (x)\right\}  .
\end{eqnarray}
The second term in the above equation is the leading twist wave function
(twist-2),
while the first and third terms are sub-leading twist (twist-3) wave functions.
If the $K^*$ meson is transversely polarized, its wave
function is then
\begin{eqnarray}
<{K^*}^-(P,\epsilon_T)|\bar{d_{\alpha}}(z)u_{\beta}(0)|0> &=&
\frac{1}{\sqrt{2N_c}}\int_0^1 dx e^{ixP\cdot z}
\left\{ \not \epsilon \left[\not p_{K^*}
\phi_{K^*}^T (x) + m_{K^*} \phi_{K^*}^v (x) \right] \nonumber\right.\\
&&\left.+ i m_{K^*} \epsilon_{\mu\nu\rho\sigma}\gamma_5\gamma^\mu
\epsilon^\nu n^\rho v^\sigma\phi_{K^*}^a (x)\right\}  .
\end{eqnarray}
 Here the leading twist wave function for the transversely
 polarized $K^*$ meson is  the first term which is proportional to $\phi_{K^*}^T $.

\section{$B\to P$ Form Factors}

The $B\to P$ form factors are defined as following:
  \begin{equation}
    \label{5}
    \langle P(p_1) |\bar q  \gamma_\mu  b|\bar  B(p_B)\rangle =\left
    [(p_B +p_1)_\mu - \frac{m_B^2-m_P^2}{q^2}q_\mu \right ]
    F_1 (q^2)
    +\frac{m_B^2-m_P^2}{q^2}q_\mu F_0 (q^2) ,
  \end{equation}
where $q=p_B - p_1$. In order to cancel the poles at $q^2 =0$,
we must impose the condition
       $$F_1(0)=F_0(0).$$
That means in the large recoil limit, we need only calculate
one independent form factor for the vector current. For the
tensor operator, there is also only one independent form factor, which is important for
the semi-leptonic decay $B\to K \ell^+ \ell^-$:
  \begin{equation}
   \label{6}
   \langle P(p_1) |\bar q  \sigma_{\mu\nu}  b|\bar  B(p_B)\rangle =i\left
   [{p_1}_\mu  q_\nu- q_\mu {p_1}_\nu \right ]
  \frac{2 F_T (q^2)}{
   m_B+m_P}  ,
   \end{equation}
\begin{equation}
 \label{7}
 \langle P(p_1) |\bar q  \sigma_{\mu\nu}\gamma_5 b|\bar  B(p_B)\rangle =
 \epsilon_{\mu \nu\alpha\beta}  {p_1}^\alpha q^\beta
\frac{2 F_T (q^2)}{
 m_B+m_P}  .
 \end{equation}

In the previous section
 we have discussed the wave functions   of
the factorization formula in eq.(\ref{factf}).
In this section, we will calculate the hard part $H$.
This part involves the current operators and the necessary hard
gluon connecting the current operator and the spectator quark.
Since the final results are expressed as integrations of the distribution
function variables, we will show the whole amplitude for each diagram
including wave functions and Sudakov factors.

There are two types of diagrams contributing to the $B\to K$ form
factors which are shown in Fig.1.
 The sum of their
amplitudes is given as
\begin{eqnarray}
F_1 (q^2=0)&=& F_0 (q^2=0)\nonumber\\
&=& 8  \pi C_F   m_B^2 \int_{0}^{1}d x_{1}d
x_{2}\,\int_{0}^{\infty} b_1d b_1 b_2d b_2\,
\left\{h_e(x_1,x_2,b_1,b_2)\left(\phi_B(x_1,b_1) \right.\right.
\nonumber \\
& & \left[(1+x_2)\phi^A_K(x_2,b_2)
 + r_K  (1-2x_2) \left(\phi^P_K(x_2,b_2) +\phi^t_K(x_2,b_2)
 \right)
 \right] -\bar \phi_B(x_1,b_1)
\nonumber \\
& & \left.\left[\phi^A_K(x_2,b_2)
 - r_K  x_2 \left(\phi^P_K(x_2,b_2) +\phi^t_K(x_2,b_2)
 \right)
 \right] \right)\alpha_s (t_e^1) \exp[-S_{ab}(t_e^1)]+\label{b}
 \\
& & \left.  2 r_K \phi^P_K(x_2,b_2)  \phi_B(x_1,b_1)
\alpha_s (t_e^2)
h_e(x_2,x_1,b_2,b_1)\exp[-S_{ab}(t_e^2)] \right \}\;, \nonumber
\end{eqnarray}
where $ r_K = m_{0K} / m_B = m_K^2 / [ m_B (m_s+m_d)]$; $C_F=4/3$
is a color factor.
The function $h_e$, the scales $t_e^i$
and the Sudakov factors $S_{ab}$ are displayed at the end
of this section.

For $B\to \pi$ form factors, one need only replace the above K
meson distribution amplitudes $\phi^i_K$  by pion distribution amplitudes $\phi^i_\pi$ and
replace the scale parameter $r_K$ by
  $ r_\pi = m_{0\pi} / m_B = m_\pi^2 / [ m_B (m_u+m_d)]$, respectively.

For the tensor operator we get the form factor formulas as
    \begin{eqnarray}
    F_T (q^2=0)&=& 8  \pi C_F   m_B^2 \int_{0}^{1}d x_{1}d
    x_{2}\,\int_{0}^{\infty} b_1d b_1 b_2d b_2\, \left\{ h_e(x_1,x_2,b_1,b_2)
    \left(\phi_B(x_1,b_1)
   \right.\right. \nonumber \\
    & &
    \left[ \phi^A_K(x_2,b_2)
     - x_2 r_K  \phi^P_K(x_2,b_2) +r_K (2+x_2)\phi^t_K(x_2,b_2)
     \right] -\bar \phi_B(x_1,b_1)\left[ \right.
    \nonumber \\
    & & \left. \left.
\phi^A_K(x_2,b_2)
     - r_K  \phi^P_K(x_2,b_2) +r_K \phi^t_K(x_2,b_2)
     \right] \right)
    \alpha_s (t_e^1)\exp[-S_{ab}(t_e^1)]
    \nonumber \\
    & & \left. + 2 r_K h_e(x_2,x_1,b_2,b_1) \alpha_s (t_e^2)   \phi_B(x_1,b_1)
    \phi^P_K(x_2,b_2)
   \exp[-S_{ab}(t_e^2)] \right \}\;.
    \label{c}
    \end{eqnarray}

In the above equations, we have used the assumption that $x_1 <<x_2$.
Since the light quark momentum fraction $x_1$ in $B$ meson is peaked at
the small region, while quark momentum fraction $x_2$
 of
K meson is peaked
around $0.5$, this is not a bad approximation.
The numerical results also show that this approximation makes very little
difference in the final result.
After using this approximation, all the diagrams are functions of
$k_1^-= x_1 m_B/\sqrt{2}$ of B meson only, independent of the variable of
$k_1^+$.

The function $h_e$, coming from the
Fourier transform of the hard amplitude $H$, is
\begin{eqnarray}
h_e(x_1,x_2,b_1,b_2)&=&
 K_{0}\left(\sqrt{x_1x_2} m_B b_1\right)
 \left[\theta(b_1-b_2)K_0\left(\sqrt{x_2} m_B
b_1\right)I_0\left(\sqrt{x_2} m_B b_2\right)\right.
\nonumber \\
& &\;\;\;\;\left.
+\theta(b_2-b_1)K_0\left(\sqrt{x_2}  m_B b_2\right)
I_0\left(\sqrt{x_2}  m_B b_1\right)\right] S_t(x_2)
\label{ha}     ,
\end{eqnarray}
where $J_0$ is the Bessel function and  $K_0$, $I_0$ are
modified Bessel functions.

The Sudakov factors used in the text are defined as
\begin{eqnarray}
S_{ab}(t) &=& s\left(x_1 m_B/\sqrt{2}, b_1\right)
+s\left(x_2 m_B/\sqrt{2}, b_2\right)
+s\left((1-x_2) m_B/\sqrt{2}, b_2\right) \nonumber \\
& &-\frac{1}{\beta_1}\left[\ln\frac{\ln(t/\Lambda)}{-\ln(b_1\Lambda)}
+\ln\frac{\ln(t/\Lambda)}{-\ln(b_2\Lambda)}\right],
\label{wp}
\end{eqnarray}
where the function $s(q,b)$ are defined in the Appendix A of ref.\cite{luy}.
The hard scale $t_i$'s in the above equations are chosen as the
largest scale of the
virtualities of internal particles in the hard $b$-quark decay
diagrams,
\begin{eqnarray}
t_{e}^1 &=& {\rm max}(\sqrt{x_2} m_B,1/b_1,1/b_2)\;,\nonumber\\
t_{e}^2 &=& {\rm max}(\sqrt{x_1}m_B,1/b_1,1/b_2)\;.
\label{scaleqg}
\end{eqnarray}

\section{$B\to V$ Form Factors}

 For the $B\to K^*$ form factors, we first define the axial vector
 current part
    \begin{eqnarray}
  \langle K^* (p_1) |\bar q \gamma_\mu  \gamma_5 b|\bar   B (p_B)  \rangle
  &=&i\left( \epsilon_\mu^* -\frac{\epsilon^*\cdot q}{q^2}q_\mu\right)
  (m_B+ m_{K^*}) A_1 (q^2) \nonumber \\
  &-&i\left( (p_B+p_1)_\mu -\frac{(m_B^2 -m_{K^*}^2)}{q^2}q_\mu\right)
  (\epsilon^* \cdot q) \frac {A_2 (q^2) }{m_B+ m_{K^*}}\nonumber \\
  &+&i\frac{2m_{K^*} (\epsilon^* \cdot q)}{ q^2} q_\mu A_0 (q^2) ,
  \label{s3}
  \end{eqnarray}
  where $\epsilon^*$  is the polarization vector
  of  $K^*$ meson.
  To cancel the poles at $q^2=0$, we must have
    \begin{equation}
    \label{s4}2m_{K^*}  A_0 (0)=
  (m_B+ m_{K^*}) A_1 (0)
  -(m_B-m_{K^*}) A_2 (0) .
  \end{equation}
For the vector current, only one form factor $V$ is defined
 \begin{eqnarray}
  \langle K^* (p_1) |\bar q \gamma_\mu b |\bar   B (p_B)  \rangle &=&
  \epsilon _{\mu\nu\alpha\beta}
  \epsilon^{\nu*} p_B^\alpha p_1^\beta \frac{2
  V(q^2)}{(m_B+m_{K^*})}.
  \end{eqnarray}
And for the tensor operators, three form factors are defined:
  \begin{eqnarray}
\langle K^* (p_1) |\bar q \sigma_{\mu\nu} b| \bar B(p_B)  \rangle &=&
-i\epsilon _{\mu\nu\alpha\beta} \epsilon^{\alpha*} p_1 ^\beta {
T_1(q^2)} - i\epsilon _{\mu\nu\alpha\beta} \epsilon^{\alpha*}
p_B^\beta {T_2(q^2)}\nonumber\\
&&-iT_3(q^2) \frac{(p_B\cdot \epsilon^*)}{p_B\cdot p_1} \epsilon_{\mu\nu\alpha\beta}
p_1^\alpha p_B^\beta, \label{ss5}
\end{eqnarray}
  \begin{eqnarray}
\langle K^* (p_1) |\bar q \sigma^{\mu\nu}\gamma_5 b|\bar   B(p_B)  \rangle &=&
 \left( p_1^\mu p_B^{*\nu}-
 p_B^{*\mu} p_1^\nu\right) \frac{(p_B \cdot \epsilon^*)}{p_B\cdot p_1}
 {T_3
(q^2) }  +  \left( \epsilon^{*\mu} p_B^\nu-
 p_B^\mu \epsilon^{*\nu}\right)   {T_2
(q^2) }\nonumber \\
&+&\left[\epsilon^{*\mu}  p_1^\nu-
   p_1^\mu \epsilon^{*\nu}\right]   {T_1
(q^2) } . \label{ss56}
\end{eqnarray}
Another frequently used   set of tensor form factors are defined
as below \cite{col}:
    \begin{eqnarray}
   \langle K^* (p_1) |\bar q \sigma_{\mu\nu}q^\nu\frac{(1+\gamma_5)}{2} b|\bar   B(p_B)
    \rangle &=& 2
   i\epsilon _{\mu\nu\alpha\beta} \epsilon^{\nu*} p_B^\alpha p_1 ^\beta {
   T_1'(q^2)}    \nonumber \\
    &+&
    \left[ \epsilon_\mu^* (m_B^2 - m_{K^*}^2) - (q \cdot \epsilon^*)
   ( p_1 +p_B)_\mu \right] {T_2' (q^2) }    \nonumber \\
  &+& (q \cdot \epsilon^*)
      \left[ q_\mu -  \frac{q^2}{  m_B^2 - m_{K^*}^2}
   (p_1+ p_B)_\mu \right]   {T_3'
   (q^2) } . \label{t'}
    \end{eqnarray}
  They are useful for the discussion of the flavor changing neutral current
  decay $B \to K^* \gamma$ and $ B\to K^* \ell^+\ell^-$.
  The relation between the two set of form factors are
  \begin{eqnarray}
  T_1'(q^2) &=& \frac{1}{4} \left[T_1(q^2) +T_2(q^2)\right]\\
 T_2'(q^2) &=&        \frac{1}{4}\left[T_1(q^2) +T_2(q^2) +\frac{q^2}{ m_B^2 -m_{K^*}^2}
  \left(T_2(q^2)
  -T_1(q^2)\right)\right]\\
 T_3'(q^2) &=&        \frac{1}{4}\left[T_1(q^2) -T_2(q^2) -\frac{m_B^2-m_{K^*}^2}{ p_B \cdot p_1 }
  T_3(q^2)\right]  .
   \end{eqnarray}
    It is easy to see from the above that at large recoil limit $q^2=0$, $T_1' (0) = T_2' (0)$.

As for $B\to \rho$, $B\to \omega$ form factors, the definition is
similar to the above, just replacing $K^*$ by $\rho$ and $\omega$
respectively.

Calculating the corresponding amplitude for Fig.1(a) and (b), we get
the formulas for the form factors at large recoil as
\begin{eqnarray}
A_0 (q^2=0)&=& 8   \pi C_F    m_B^2 \int_{0}^{1}d
x_{1}d x_{2}\,\int_{0}^{\infty} b_1d b_1 b_2d b_2\,
\times \left\{ \alpha_s(t_e^1)\exp[-S_{ab}(t_e^1)] \left[\phi_B(x_1,b_1)\right.\right.
\nonumber \\
& &\left((1+x_2) \phi_{K^*} (x_2, b_2)
+(1-2x_2)r_{K^*} \left(\phi_{K^*}^t (x_2, b_2)+\phi_{K^*}^s (x_2, b_2)
\right)
\right) -\bar\phi_B(x_1,b_1)\nonumber \\
&&\left.
\left(\phi_{K^*}(x_2, b_2)-x_2 r_{K^*} \left(\phi_{K^*}^t (x_2, b_2)+\phi_{K^*}^s (x_2, b_2)
\right)\right)\right] h_e(x_1,x_2,b_1,b_2)+ \label{aa} \\
&&\left.2 r_{K^*}  \phi_B(x_1,b_1)  \phi_{K^*}^s (x_2, b_2)
\alpha_s(t_e^2) h_e(x_2,x_1,b_2,b_1)\exp[-S_{ab}(t_e^2)] \right\}\;,
 \nonumber
\end{eqnarray}
where $r_{K^*}=m_{K^*}/m_B$. The form factor $A_0$ is the one
contributing to the non-leptonic B decays $B \to P V$, where the
vector meson is longitudinally polarized.     This is shown in the
above equation (\ref{aa}) that the formula depends only on longitudinal wave
functions.
  On the other hand,
the form factor $A_1$ contributing to  the $B\to V V$ decays
depends only on transverse wave functions, which is shown below
\begin{eqnarray}
A_1 (q^2=0)&=& 8   \pi C_F    m_B(m_B-m_{K^*}) \int_{0}^{1}d
x_{1}d x_{2}\,\int_{0}^{\infty} b_1d b_1 b_2d b_2\, \left\{  h_e(x_1,x_2,b_1,b_2)
\exp[-S_{ab}(t_e^1)]\right.
\nonumber \\
& & \alpha_s(t_e^1)\left[
\phi_B(x_1,b_1) \left(\phi_{K^*}^T (x_2, b_2)
+(2+x_2)r_{K^*} \phi_{K^*}^v (x_2, b_2)-r_{K^*}x_2\phi_{K^*}^a (x_2, b_2)
\right) \right. \nonumber \\
&&\left.-\bar\phi_B(x_1,b_1)\left(\phi_{K^*}^T (x_2, b_2)
+r_{K^*} \phi_{K^*}^v (x_2, b_2)-r_{K^*}  \phi_{K^*}^a (x_2, b_2)
\right)\right]+ r_{K^*} \label{ab}\\
&&\left.\phi_B(x_1,b_1)  [\phi_{K^*}^v (x_2, b_2)+\phi_{K^*}^a (x_2,
b_2)]
\alpha_s(t_e^2) h_e(x_2,x_1,b_2,b_1)\exp[-S_{ab}(t_e^2)] \right\}\;.
\nonumber
\end{eqnarray}
The form factor $A_2$ can be calculated from eq.(\ref{s4}), using
the above eqs.(\ref{aa}, \ref{ab}) for $A_0$ and $A_1$.   It
depends on both transverse and longitudinal wave functions.

The vector form factor $V$ depending  only on transverse wave functions, is expressed as
\begin{eqnarray}
V (q^2=0)&=& 8   \pi C_F    m_B(m_B+m_{K^*}) \int_{0}^{1}d
x_{1}d x_{2}\,\int_{0}^{\infty} b_1d b_1 b_2d b_2\,  \left\{
\alpha_s(t_e^1) h_e(x_1,x_2,b_1,b_2)\right.
\nonumber \\
& & \left(\phi_B(x_1,b_1)\left[ \phi_{K^*}^T (x_2, b_2)
+(2+x_2)r_{K^*} \phi_{K^*}^a (x_2, b_2)-r_{K^*}x_2\phi_{K^*}^v (x_2, b_2)
\right] -  \right.\nonumber \\
&&\left.\bar \phi_B(x_1,b_1) \left[ \phi_{K^*}^T (x_2, b_2)
+r_{K^*} \phi_{K^*}^a (x_2, b_2)-r_{K^*} \phi_{K^*}^v (x_2, b_2)
\right] \right) \exp[-S_{ab}(t_e^1)] \nonumber \\
&&+ r_{K^*}h_e(x_2,x_1,b_2,b_1)  \left[\phi_{K^*}^v (x_2, b_2)+\phi_{K^*}^a (x_2,
b_2)\right] \phi_B(x_1,b_1)\;\\
&& \left.\alpha_s(t_e^2) \exp[-S_{ab}(t_e^2)] \right\}\;.
\nonumber
\label{ac}
\end{eqnarray}

As for the tensor form factors, $T_1$ and $T_2$ contributing to the $B\to K^*\gamma$ decay,
depend only on transverse wave functions
\begin{eqnarray}
T_1 (q^2=0)&=& 16   \pi C_F    m_B^2 \int_{0}^{1}d
x_{1}d x_{2}\,\int_{0}^{\infty} b_1d b_1 b_2d b_2\,  \left\{
\alpha_s(t_e^1) h_e(x_1,x_2,b_1,b_2)\left(
\phi_B(x_1,b_1) \right.\right.
\nonumber \\
& & \left[ (1+x_2) \phi_{K^*}^T (x_2, b_2)
+2(1-x_2)r_{K^*} \phi_{K^*}^a (x_2, b_2)-2x_2 r_{K^*} \phi_{K^*}^v (x_2, b_2)
\right] - \nonumber \\
&&\left.\bar \phi_B(x_1,b_1) \left[ \phi_{K^*}^T (x_2, b_2)
+(1-x_2) r_{K^*} \phi_{K^*}^a (x_2, b_2)-(1+x_2) r_{K^*} \phi_{K^*}^v (x_2, b_2)
\right] \right)  \nonumber \\
&&\times \exp[-S_{ab}(t_e^1)]
+ r_{K^*}h_e(x_2,x_1,b_2,b_1)  \left[\phi_{K^*}^v (x_2, b_2)+\phi_{K^*}^a (x_2,
b_2)\right]
\;\label{t1}
\\
&& \left.  \times  \phi_B(x_1,b_1)
\alpha_s(t_e^2) \exp[-S_{ab}(t_e^2)] \right\}
\;.
\nonumber
\end{eqnarray}
\begin{eqnarray}
T_2 (q^2=0)&=& 16   \pi C_F    m_B^2 \int_{0}^{1}d
x_{1}d x_{2}\,\int_{0}^{\infty} b_1d b_1 b_2d b_2\,
\alpha_s(t_e^1) h_e(x_1,x_2,b_1,b_2)    r_{K^*}
\nonumber \\
& &
  \left[ \phi_{K^*}^v (x_2, b_2)- \phi_{K^*}^a (x_2, b_2)
\right]
\left(     \phi_B(x_1,b_1) -\bar \phi_B(x_1,b_1)
\right)
  \exp[-S_{ab}(t_e^1)]
\;\label{t2}
\;.
\end{eqnarray}
While form factor $T_3$ depends on both longitudinal and
transverse wave functions
\begin{eqnarray}
 T_3 (q^2=0)&=& 16   \pi C_F    m_B^2 \int_{0}^{1}d
 x_{1}d x_{2}\,\int_{0}^{\infty} b_1d b_1 b_2d b_2\,  \left\{
 \alpha_s(t_e^1) h_e(x_1,x_2,b_1,b_2)\left(
 \phi_B(x_1,b_1) \right.\right.
 \nonumber \\
 & & \left[  \phi_{K^*}  (x_2, b_2)
 +(2+x_2)r_{K^*} \phi_{K^*}^t (x_2, b_2)-x_2 r_{K^*} \phi_{K^*}^s (x_2, b_2)
 \right] - \nonumber \\
 &&\left.\bar \phi_B(x_1,b_1) \left[ \phi_{K^*} (x_2, b_2)
 + r_{K^*} \phi_{K^*}^t (x_2, b_2)- r_{K^*} \phi_{K^*}^s (x_2, b_2)
 \right] \right)  \nonumber \\
 &&\times \exp[-S_{ab}(t_e^1)]
 +2 r_{K^*}h_e(x_2,x_1,b_2,b_1)  \phi_{K^*}^s (x_2, b_2)
 \;\label{t3}
 \\
 && \left.  \times  \phi_B(x_1,b_1)
 \alpha_s(t_e^2) \exp[-S_{ab}(t_e^2)] \right\}r_{K^*}   -2r_{K^*}^2 T_1
 -T_2
 \;.
 \nonumber
 \end{eqnarray}

\section{Numerical Calculations and Discussions}

In the numerical calculations we use
\begin{eqnarray}
 \Lambda_{\overline{\mathrm{MS}}}^{(f=4)} = 250 { MeV},
 & f_\pi = 132 { MeV}, &f_K = 160 { MeV},
 \nonumber\\
      f_B = 190 MeV,  &
 m_{0\pi} = 1.4 { GeV},& m_{0K} = 1.7 { GeV}, \nonumber\\
   M_B = 5.2792 { GeV}, &f_{K^*} = 220
 { MeV}, &f_{K^*}^T = 180  MeV, \nonumber\\
     M_W = 80.41{ GeV},&f_{\rho} = 217
  { MeV}, &f_{\rho}^T = 160  MeV, \nonumber\\
  &    f_{\omega} = 195
  { MeV}, &f_{\omega}^T = 160  MeV.
  \label{para}
\end{eqnarray}
The distribution amplitudes $\phi_\pi^i(x)$,  $\phi_K^i(x)$, $\phi_\rho^i(x)$ ($\phi_\omega^i(x)$) and
  $\phi_{K^*}^i(x)$  of the light mesons
used in the numerical calculation are listed in Appendix A.

\begin{figure}[ht]
\begin{center}
\vspace{1cm}
\scalebox{1} {
\epsfig{file=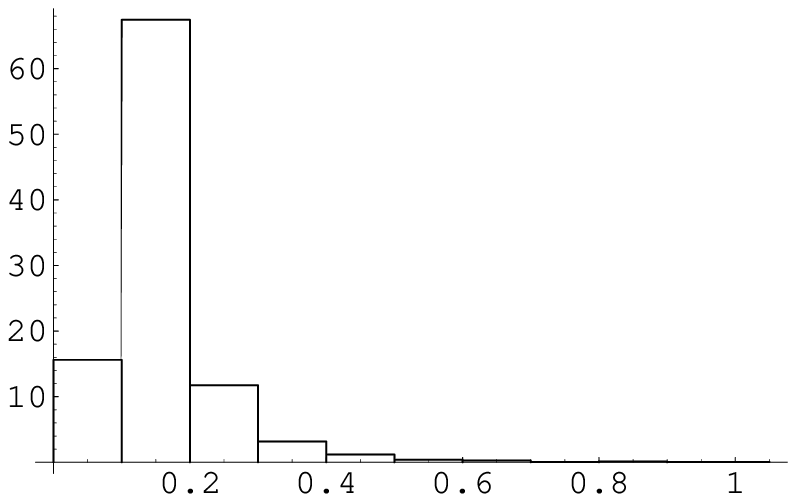}
\begin{picture}(0,0)(0,0)
   \put(-30,-10){$\alpha_s/\pi$ }
   \put(-225,150){$\%$ }
   \put(-120,-30){(a)}
  \end{picture} }
\scalebox{0.9} {
\epsfig{file=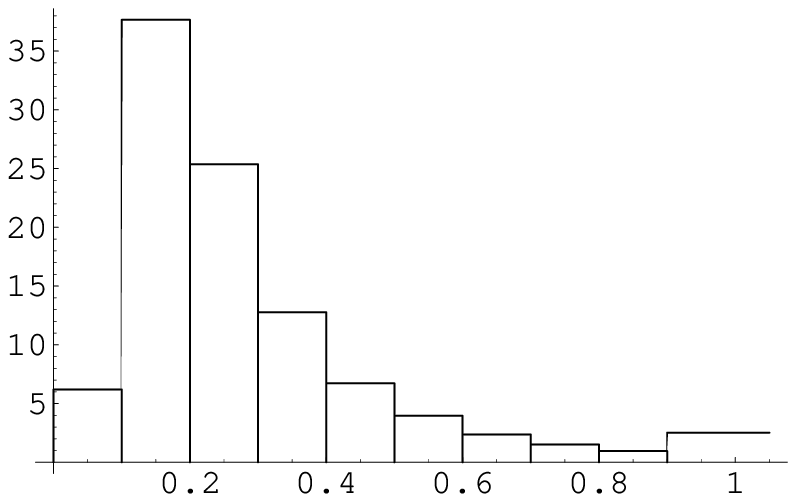}
\begin{picture}(0,0)(0,0)
   \put(-30,-10){$\alpha_s/\pi$ }
   \put(-225,150){$\%$ }
   \put(-120,-30){(b)}
  \end{picture} }
\end{center}
\caption{Contributions to $B\to \pi$ transition form factors
(\small{$F^{B\pi}(0)$})  from different ranges of $\alpha_s/\pi$,
(a) with the hard scale chosen as virtualities of internal particles including both
quarks and gluons; (b) the hard scale chosen as virtualities of only internal gluons. }
\label{00}
\end{figure}

Fig.\ref{00}(a) displays the contributions to $B \to \pi$
transition form factor at the large recoil limit
$q^2\to 0$ from different ranges of $\alpha_s/\pi$, where
the hard scale $t$ is chosen as eq.(\ref{scaleqg}), i.e.,
the maximum virtuality of both internal quarks and gluons
 in the hard $b$-quark decay diagrams. It shows that most
of the contribution comes from the range $\alpha_s/\pi<0.3$,
implying that the average scale is around
$\sqrt{\Lambda_{QCD}m_B}$.
For other $B$ to light meson transition form factor calculations,
we have very similar results. It is observed that with the hard
scale chosen in eq.(\ref{scaleqg}), PQCD is applicable to
$B\to\mbox{light~meson}$ transition form factors. Recent study
shows that PQCD is even applicable to $B\to D^{(*)}$ form factors
\cite{kurilisanda}. However, different perturbative percentage distribution
over $\alpha_s/\pi$ was observed in \cite{weiyang,dgsh}.
We check the reason which causes this difference and find
that the most important reason is the way of choosing the
hard scale $t$. If the hard scale is chosen as the maximum
virtuality of only the gluon and other transverse momentum scales, i.e.,
$t\equiv max(\sqrt{x_1x_2}m_B,1/b_1,1/b_2)$, the perturbative
percentage distribution will be similar to ref.\cite{weiyang,dgsh},
as shown in
Fig.\ref{00}(b).
 Therefore
the way of choosing the hard scale is one of the important
ingredients in the PQCD approach, which deserves more concern.
\footnote{By numerical check we find that
these two different choices of the hard scale only
slightly affect the magnitude of the form factors. For example
it can only change the $B\to\pi$ from factor by a few percent.}
Provided that the virtuality of the internal quark momentum
(longitudinal) must appear as a characteristic scale in the
hard diagram, in general it should be taken into account.
Therefore we think that it is reasonable to choose
both the virtualities of internal quarks and gluons as
the hard scale. Certainly the most powerful proof of this point
should be performed under the help of numerical calculation
of higher order loop corrections. However, such a deeper
discussion of this problem is beyond the scope
of this paper, which shall be left to other attempts.

\begin{table}[htb]
\caption{ $B$ meson transition form factors at $q^2=0$ with
  the hard scale chosen in eq.(\ref{scaleqg}), and the numbers
  in parentheses are results without the contribution of
  $\bar{\phi}_B$ .}
 \doublerulesep 1.2pt \tabcolsep 0.1cm
 \begin{center} {\small
\begin{tabular}{c|ccccccc}
\hline \hline\
process & $F_0(0)=F_1(0)$ & $F_T(0)$ &
  &  &   &  &  \\
\hline\hline
$B\to\pi$ & $0.292\pm 0.030$  & $0.278\pm 0.028$  & &  & & & \\
          & (0.199)& (0.189) & & & & & \\ \hline
$B\to K$ & $0.321\pm 0.036$ & $0.311\pm 0.033$ & & & & & \\
      &(0.231)&(0.223)& & & &  & \\  \hline  \hline
 process & $V(0)$ & $A_0(0)$ &
 $A_1(0)$ & $A_2(0)$ & $T_1(0)$ & $T_2(0)$ &$T_3(0)$ \\
 \hline\hline
$B\to\rho$  & $0.318\pm 0.032$ & $0.366\pm 0.036$ & $0.25\pm 0.02$ & $0.21\pm 0.01$ &
$0.56 \pm 0.05$
   & $0.013\pm 0.001$ & $0.06 \pm 0.01$\\
            &(0.226)&(0.256)&(0.17)&(0.14)&(0.41)
   &(0.004)&(0.05) \\   \hline
$B\to\omega$&  $0.305\pm 0.030$ & $0.347\pm 0.036$ & $0.24\pm 0.02$
& $0.20\pm 0.02$ & $0.53\pm 0.05$ & $0.012
\pm 0.001$ &   $0.06 \pm 0.01$ \\
   &(0.212)&(0.250)&(0.16)&(0.13)&(0.38)&(0.003)&
  (0.05)\\ \hline
$B\to K^{*}$  & $0.406\pm 0.042$ & $0.455\pm 0.047$ & $0.30\pm 0.03$ & $0.24\pm 0.02$
& $0.69\pm 0.08$ & $0.007\pm 0.001$ &
  $0.09 \pm 0.01$ \\
  &(0.293)&(0.336)&(0.21)&(0.16)&(0.51)&(-0.001) &
  (0.07) \\ \hline          \hline
\end{tabular} }
\end{center}
\label{tt1}
\end{table}

The results of $B\to P,~V$ light meson transition form factors are
given in Table \ref{tt1} with the hard scale chosen in
eq.(\ref{scaleqg}). Compared with previous PQCD calculations on
some
$B\to P,~V$ transition form factors
\cite{kurimoto,kurilisanda,chengeng}, the current work is
different from them mainly on two points:
\begin{enumerate}
\item The $B$ meson wave function
used here is what derived from the equation of motion in Heavy
Quark Effective Theory \cite{kkqt}. There is only one free
parameter in the functions of distribution amplitudes,
$\bar\Lambda$. We show the results for $\bar \Lambda =(700 \pm 50
)$MeV in table~\ref{tt1}. All the form factors are sensitive to
this parameter, i.e. sensitive to the shape of the B meson
distribution amplitudes.
\item  Two Lorentz structure
terms of $B$ meson wave function, both $\phi_B$ and $\bar{\phi}_B$
defined in eqs.(\ref{aa1}, \ref{phiB}), are taken into account in this work.
To see how large the $\bar{\phi}_B$ term contributes, we give
the results without the contribution of $\bar{\phi}_B$
in the parentheses in Table 1. They show that the contribution of
$\bar{\phi}_B$ is about $30\%$. The dominant contribution comes
from $\phi_B$ term.
This result shows that simply dropping the contribution of $\bar \phi_B$
is not a good approximation.
 \end{enumerate}

\begin{table}[ht]
\caption{Form factors at $q^2=0$ for $B\to \pi$ and $B\to\rho$
  transitions calculated in this work and UKQCD .}
 \doublerulesep 1.2pt \tabcolsep 0.1cm
 \begin{center} {\small
\begin{tabular}{c|ccccccc}
\hline \hline
       & $F_0(0)=F_1(0)$ & $V(0)$ & $A_0(0)$ &$A_1(0)$ & $A_2(0)$ \\
\hline\hline
UKQCD \cite{ukqcd} & $0.27\pm 0.11$ & $0.35^{+0.06}_{-0.05}$ &
       $0.30^{+0.06}_{-0.04}$ & $0.27^{+0.05}_{-0.04}$ &
       $0.26^{+0.05}_{-0.03}$ \\ \hline
this work & 0.292  & 0.318 & 0.366 & 0.250 & 0.210 \\   \hline
\hline
\end{tabular} }
\end{center}
\label{tt2}
\end{table}

We compare some of the results calculated in this work with lattice
calculation by UKQCD collaboration \cite{ukqcd} in Table 2. It
shows that our results are consistent with theirs.

The $B\to K^*$ form factors are useful for the calculation of
flavor changing neutral current process $B\to K^* \gamma$ and
$B\to K^* \ell^+\ell^-$, which have been discussed many times \cite{rev}.
We show some of them in table~\ref{tt3} for comparison.
 It is easy to see that our results agree with lattice calculations \cite{ukqcd} and the results
 calculated using lattice-constrained dispersion quark model \cite{gi}.

\begin{table}[ht]
\caption{Some form factors at $q^2=0$ for $B\to K^*$
  transitions calculated in this work and some other works.}
 \doublerulesep 1.2pt \tabcolsep 0.1cm
 \begin{center} {\small
\begin{tabular}{c|ccc}
\hline \hline\
       & $T_1'(0)=T_2'(0)$ & $A_0(0)$ &$A_1(0)$  \\
\hline\hline
quark model \cite{qm} & $0.155$ & $0.32$ &
       $0.26$             \\ \hline
 QCD sum rule \cite{col} & $0.19\pm 0.03$ & $0.3\pm 0.03$ &
         $0.37\pm 0.03$             \\ \hline
light cone sum rule \cite{lcsr} & $0.18$ & $0.27$ &
        $0.36$             \\ \hline
 Lattice        QCD  \cite{ukqcd} & $0.16^{+0.02}_{-0.01}$ & $0.33$ &
                 $0.29$             \\ \hline
 dispersion quark model \cite{gi} & $0.177$ & $0.44$ &
         $0.33$             \\ \hline
this work & 0.175  & 0.455 & 0.297 \\   \hline \hline
\end{tabular} }
\end{center}
\label{tt3}
\end{table}

\section{Summary}

We have calculated $B\to P$ and $B\to V$ transition form factors
in the PQCD approach. We not only calculate the $B$ to light meson
transition form factors defined in vector and axial vector
currents, but also form factors defined in tensor currents
$\bar{q}\sigma_{\mu\nu} b$ and $\bar{q}\sigma_{\mu\nu}\gamma_5 b$,
which can be used to study semi-leptonic and radiative $B$ decays induced by
magnetic penguin operators $\bar{q}\sigma_{\mu\nu}(1+\gamma_5) b
F_{\mu\nu}$. With the hard scale appropriately chosen, Sudakov
effects can effectively suppress long-distance dynamics, which
makes short-distance contribution dominate. The characteristic
scale in $B$ to light meson transition processes is around
$\sqrt{\Lambda_{QCD} m_B}$.

We considered both of the two Lorentz
structures of $B$ meson wave functions, and found that the
contribution of $\bar{\phi}_B$ defined in eqs.(\ref{aa1}, \ref{phiB})
is about 30\%.

Finally we compared our results with Lattice calculation, some quark model and
QCD sum rule calculations,
we found that  they are consistent with our results.

\section*{Acknowledgments}
We thank T. Kurimoto, H.n. Li,  S. Mishima and D. Pirjol for reading the
manuscript and making helpful comments.
This work is supported in part by National  
Science Foundation of China with contract No.10205017,
 90103013 and 10135060.

\begin{appendix}

\section{Wave Functions of Light Mesons Used in the Numerical Calculation}

 For the light meson wave function, we neglect the $b$ dependence part, which is not
 important in numerical analysis.

 The distribution amplitude $\phi_\pi^A$ for the twist-2 wave function  and the distribution
 amplitudes $\phi_\pi^P$ and $\phi_\pi^t$ of
 twist-3 wave functions  are taken
  from \cite{ball},
 \begin{eqnarray}
 \phi_\pi^A(x) &=&  \frac{3 f_\pi}{\sqrt{6} }
   x (1-x)  \left[1+0.44C_2^{3/2} (t) +0.25 C_4^{3/2}
  (t)\right],\label{piw1}\\
 \phi_\pi^P(x) &=&   \frac{f_\pi}{2\sqrt{6} }
   \left[ 1+0.43 C_2^{1/2} (t) +0.09 C_4^{1/2} (t) \right]  ,\\
 \phi_\pi^t(x) &=&  \frac{f_\pi}{2\sqrt{6} } ~t
   \left[ 1+0.55  (10x^2-10x+1)  \right]  ,   \label{piw}
 \end{eqnarray}
where $t=1-2x$.  The Gegenbauer polynomials are defined by
 \begin{equation}
 \begin{array}{ll}
 C_2^{1/2} (t) = \frac{1}{2} (3t^2-1), & C_4^{1/2} (t) = \frac{1}{8}
 (35t^4-30t^2+3),\\
 C_2^{3/2} (t) = \frac{3}{2} (5t^2-1), & C_4^{3/2} (t) = \frac{15}{8}
 (21t^4-14t^2+1),
 \end{array}
 \end{equation}
whose coefficients correspond to $m_{0\pi}=1.4$ GeV.

 We choose the different distribution amplitudes  of ${\rho}$ meson longitudinal wave function as \cite{ball2},
\begin{eqnarray}
\phi_{\rho}(x) &=&  \frac{3f_{\rho}}{\sqrt{6} }
   x (1-x)  \left[1+ 0.18C_2^{3/2} (t) \right],
 \\
    \phi_{\rho}^t(x) &=&  \frac{f_{\rho}^T }{2\sqrt{6} }
  \left\{  3 t^2 +0.3t^2  \left[5t^2-3  \right]
 +0.21 \left[3- 30 t^2 +35 t^4\right] \right\},\\
\phi_{\rho}^s(x) &=&  \frac{3f_{\rho}^T  }{2\sqrt{6} }
 ~ t  \left[1+ 0.76 (10 x^2 -10 x +1) \right] .
\end{eqnarray}

    For the transverse $\rho$ meson  we use \cite{ball2}:
    \begin{eqnarray}
\phi_{\rho}^T(x) &=&  \frac{3f_{\rho}^T }{\sqrt{6} }
  x (1-x)  \left[1+ 0.2C_2^{3/2} (t) \right],
 \\
    \phi_{\rho}^v(x) &=&  \frac{f_{\rho} }{2\sqrt{6} }
  \left\{  \frac{3}{4} ( 1+ t^2) +0.24( 3 t^2-1) +0.12 ( 3-30 t^2
  +35 t^4) \right\},\\
\phi_{\rho}^a(x) &=&  \frac{3 f_{\rho}}{4\sqrt{6}
 }~  t  \left[1+ 0.93 (10 x^2 -10 x +1) \right] .
\end{eqnarray}
  For the $\omega$ meson, we use the same as the above $\rho$
  meson, except changing the decay constant $f_\rho$ with $f_\omega$.

 We use   $\phi_K^A$ of the K meson twist-2 wave function  and $\phi_K^P$ and $\phi_K^t$
 of the twist-3 wave functions
 from \cite{ball,ball2,ball3},
 \begin{eqnarray}
 \phi_K^A(x) &=&  \frac{3 f_K }{\sqrt{6} }
  x (1-x)  \left[1+0.51 t +0.3
 \{5 t^2-1\}\right],\label{Kw1}
 \\
 \phi_K^P(x) &=&   \frac{f_K}{2\sqrt{6} }
   \left[ 1+0.12 (3t^2-1) -0.12   (3-30t^2+35t^4)/8 \right]
   ,\\
 \phi_K^t(x) &=&  \frac{f_K}{2\sqrt{6} } ~t
   \left[ 1+0.35  (10x^2-10x+1)  \right]  ,    \label{Kw}
 \end{eqnarray}
 whose coefficients correspond to $m_{0K}=1.7$ GeV.

 We choose the light cone distribution amplitudes of ${K^*}$ meson longitudinal wave function  as \cite{ball2},
\begin{eqnarray}
\phi_{K^*}(x) &=&  \frac{3 f_{K^*}}{\sqrt{6} }
  x (1-x)  \left[1+ 0.57t + 0.07C_2^{3/2} (t) \right],
 \\
    \phi_{K^*}^t(x) &=&  \frac{f_{K^*}^T }{2\sqrt{6} }
  \left\{  0.3 t (3t^2 +10 t-1) +1.68 C_4^{1/2} (t)
  +0.06t^2(5t^2-3)   \right.    \nonumber\\
&& \left. +0.36 [1 -2 t -2t \ln(1- x) ] \right\},
\\
\phi_{K^*}^s(x) &=&  \frac{ f_{K^*}^T }{2\sqrt{6} }
\left\{ 3 t  \left[1+ 0.2t+0.6 (10 x^2 -10 x +1)\right] -0.12x(1-x)     \right.    \nonumber\\
&& \left. +0.36 [1 -6x-2 \ln(1- x)] \right\}.
\end{eqnarray}
The light cone distribution amplitudes of $K^*$ transverse   wave function
are used as
    \begin{eqnarray}
\phi_{K^*}^T(x) &=&  \frac{3 f_{K^*}^T }{\sqrt{6} }
 x (1-x)  \left[1+ 0.6t+0.04C_2^{3/2} (t) \right],
 \\
    \phi_{K^*}^v(x)
          &=&  \frac{f_{K^*}}{2\sqrt{6} }
      \left\{ \frac{3}{4}  (1+t^2 +0.44 t^3) +0.4 C_2^{1/2} (t)
      +0.88C_4^{1/2}(t)     \right.    \nonumber\\
    && \left. +0.48 [2x + \ln(1- x)] \right\},
    \\
\phi_{K^*}^a(x)
     &=&  \frac{ f_{K^*} }{4\sqrt{6} }
    \left\{ 3 t  \left[1+ 0.19t+0.81 (10 x^2 -10 x +1)\right] -1.14 x(1-x)     \right.
        \nonumber\\
    && \left. +0.48 [1 -6x-2 \ln(1- x)] \right\}.
\end{eqnarray}

\end{appendix}

\end{document}